\documentclass[aps,pra,twocolumn,a4paper,floatfix,
superscriptaddress, showpacs,preprintnumbers] {revtex4-1}
\usepackage{graphicx}
\usepackage{calc}
\usepackage{amsmath}
\usepackage{amssymb}

\newcommand{\D}{\mbox{\rm d}}
\newcommand{\I}{{\rm{I}}}
\newcommand{\Tr}{\mbox{\rm Tr}}

\newcommand{\ket}[1]{\left|\mathrm{#1}\right\rangle}
\newcommand{\bra}[1]{\left\langle\mathrm{#1}\right|}

\begin{document}

\title{Bell nonlocality in the turbulent atmosphere}
\preprint{PHYSICAL REVIEW A {\bf 94}, 053801 (2016)}

\author{M.O. Gumberidze}

\affiliation{Institute of Physics, National Academy of Sciences of
Ukraine,
Prospect Nauky 46, 03028 Kiev, Ukraine}%
\affiliation{Physics Department, Taras Shevchenko National University of Kiev,
Prospect Glushkova 2, 03022 Kiev, Ukraine}
\affiliation{Department of Optics, Palack\'y University, 17. listopadu 12, 771 46 Olomouc, Czech Republic}

\author{A.A. Semenov}
% \email[E-mail address: ]{sem@iop.kiev.ua}%
\affiliation{Institute of Physics, National Academy of Sciences of
Ukraine,
Prospect Nauky 46, 03028 Kiev, Ukraine}%
\affiliation{Institut f\"ur Physik, Universit\"{a}t Rostock, Albert-Einstein-Stra\ss{}e 23, D-18059
Rostock, Germany}

\author{D. Vasylyev}
\affiliation{Institut f\"ur Physik, Universit\"{a}t Rostock, Albert-Einstein-Stra\ss{}e 23, D-18059
	Rostock, Germany}
\affiliation{Bogolyubov Institute for Theoretical Physics, NAS of Ukraine, Vulytsya
	Metrologichna 14-b, 03680 Kiev, Ukraine}

\author{W. Vogel}
\affiliation{Institut f\"ur Physik, Universit\"{a}t Rostock, Albert-Einstein-Stra\ss{}e 23, D-18059
	Rostock, Germany}

\begin{abstract}
Violations of Bell inequalities are better preserved by turbulent atmospheric channels than by comparable
optical fibers in the scenario of copropagating entangled photons [A.~A.~Semenov and W.~Vogel, Phys. Rev. A \textbf{81},
023835 (2010); arXiv:0909.2492]. Here we reexamine this result for the
case of counterpropagation also considering the fact that each receiver
registers so-called double-click events, which are caused by dark counts, stray
light, and multi-photon entangled pairs. We show that advantages of the 
atmospheric links are feasible only for the copropagation scenario in the case of strong 
fluctuations of losses.
For counterpropagation, the violations of Bell inequalities can be improved with an additional postselection procedure testing the channel transmittance.

\end{abstract}


\maketitle
\section{Introduction}
\label{Introduction}

The interest in quantum key distribution 
(QKD) schemes~\cite{Takesue}
through free-space channels is due to the intriguing practical
perspectives of information security in the scenarios of 
communication between mobile participants, links through hardly
accessible regions, global quantum communications via
satellites~\cite{Satellite0,Satellite2,Satellite3,Satellite4,Satellite5,Satellite6}, etc. 
QKD protocols using Bell 
inequalities (for a review see Ref.~\cite{BellReview}) assume sharing of 
radiation-field modes between remote participants.
Nonclassical fields may violate the Bell inequalities. This corresponds to the 
absence of locality and/or realism (local realism) in quantum physics. We refer to this phenomenon, 
in accordance with \cite{BellReview}, as \textit{Bell nonlocality}.
In this context, the original Ekert protocol (E91)
\cite{Ekert} utilizes the Bell inequalities to test for eavesdropping. Moreover, the relevance 
of Bell inequalities has been recognized in the context of device-independent QKD 
(cf.~Refs.~\cite{Acin1,Acin2}). The corresponding protocols do not depend on detailed 
characterizations of measurement devices, and eavesdroppers may even have some 
control of them.

Experimental violations of Bell inequalities for light passing through the atmosphere
have been demonstrated for a 144~km channel on the Canary Islands 
\cite{Ursin, Fedrizzi}.
A consistent theoretical analysis of such types of experiments requires a deep understanding of destructive phenomena.  This includes generation, transmission, and
detection of nonclassical light. Fading effects, i.e., fluctuating losses
(cf.~Refs~\cite{Semenov2009, Vasylyev2012}), absorption, and noise events originating
from dark counts or stray light are serious obstacles for proper tests of the Bell inequalities.

In a Bell-like experiment implemented in a 144~km atmospheric
channel~\cite{Fedrizzi}, the scenario of copropagation has been
studied. In that case two photons, which were prepared at the parametric down-conversion (PDC) source in a polarization-entangled state, were sent in the same direction
from the transmitter to the receiver with a temporal separation much
smaller than the characteristic time scales of atmospheric variations (see Refs.~\cite{Avetisyan,Zhang} for other types of entangled states prepared by the PDC source, which are useful for atmospheric communications). Such fading channels can be
considered to have correlated transmittances. For this scenario the turbulence may
even improve measured values of the Bell parameter~\cite{Semenov2010}. This effect can
be easily explained by the fact that correlated counts from the source are more likely to be detected when the
channel is randomly transparent. On the other hand, for events with low transmittances the detection of simultaneous  noise clicks occurs with lower probabilities.

Multi-photon pairs from PDC sources~\cite{Ma,
Kok} and noise counts also lead to the appearance of so-called double-click
events, which make it impossible to ascribe a definite value of the qubit for the
corresponding measurements~\cite{Semenov2011}. The exclusion of such events from 
consideration may lead to a lower security of QKD protocols. Instead, one
assigns a random value to the corresponding qubit. This technique, usually referred
to as the squash model~\cite{Beaudry,Moroder,Fung}, enables one to perform a consistent
mapping of continuous-variable PDC states onto discrete-variable
qubit states.

In the present paper we consider violations of Bell inequalities caused by light being in a 
polarization-entangled state, which is transmitted through the atmosphere, and address 
two main issues. First, we study how the incorporation of double-click events in the framework of the corresponding squash models influences the feasibility of 
checking the Bell-inequality violation in turbulent-atmosphere channels. Second, we demonstrate that for the counterpropagation scenario the advantages of atmospheric channels are not directly feasible. However, the advantages can still be utilized by applying a postselection procedure testing the channel transmittance. 

The paper is organized as follows. In Sec.~\ref{Squashing} we derive
relations for the correlation coefficients and Bell parameters in the 
presence of double-click events and for scenarios of copropagating and 
counterpropagating fields. In Sec.~\ref{Copropagation} we analyze  the effect of 
double-click events on
the Bell-inequality violation in the case of copropagation. The case of
counterpropagation is considered in Sec.~\ref{Counterpropagation}. 
In Sec.~\ref{Sec:Post-processing} we consider a postselection procedure, which 
enables us to improve the violations of Bell inequalities.
In Sec.~\ref{SummaryAndConclusions} we summarize our results and give some conclusions.

\section{Bell inequalities for atmospheric channels}
\label{Squashing}

\subsection{Design of the experiment}

\begin{figure}[ht!]
\includegraphics[clip=,width=1\linewidth]{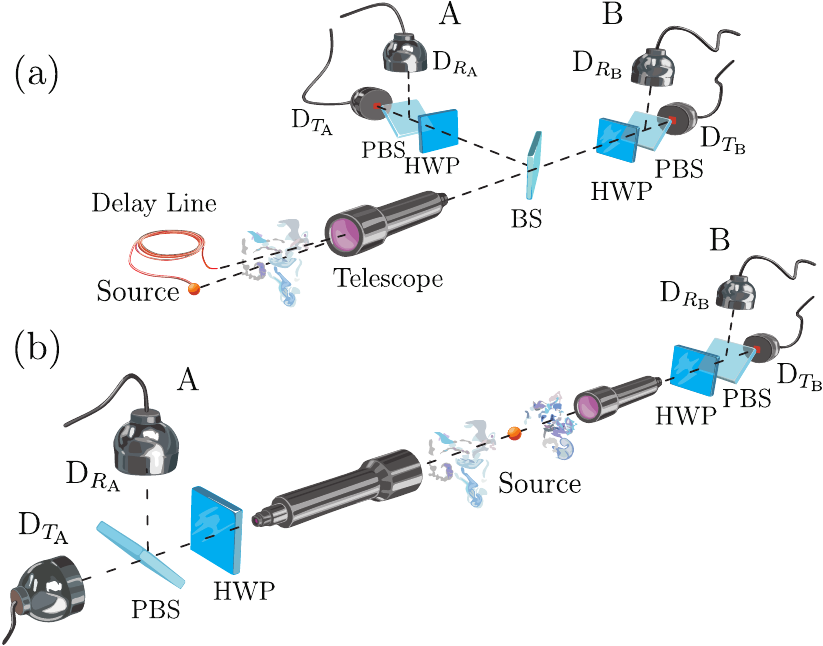}
\caption{\label{Scheme} (Color online) Typical experimental setups for 
verification of Bell inequalities in atmospheric channels: (a) the copropagation scheme as 
used in Ref.~\cite{Fedrizzi}; (b) the counterpropagation scheme. For more details, see 
the text.}
\end{figure}

Let us recall the physical background of Bell-inequality tests in atmospheric
channels~\cite{Fedrizzi,Ursin,Semenov2010}. We distinguish between two scenarios: 
the first one with different modes of the entangled light copropagating in the same 
direction and the second one with the modes being counterpropagating (cf. Fig.~\ref{Scheme}). 
The PDC source generates entangled photons, which are
sent through the turbulent atmosphere to receivers $\textrm{A}$ and $\textrm{B}$. In
the copropagation scenario one photon is sent directly and the other one through a delay
line  to the corresponding receiver. There the photons are analyzed with polarization
analyzers. Each of them consists of a half-wave plate (HWP), rotating the
polarization by the angles $\theta_\mathrm{A}$ and $\theta_\mathrm{B}$,
polarizing beam splitters (PBS), and click detectors for the
transmission and reflection channels, $\textrm{D}_{T_\mathrm{{A(B)}}}$ and
$\textrm{D}_{R_\mathrm{A(B)}}$, respectively. The scheme of copropagation additionally
includes a 50:50 beam splitter (BS), which randomly selects photons 
propagating to receivers A and B with the unavoidable introduction of
3-dB deterministic losses (cf.~Ref.~\cite{Fedrizzi}). In the scenario of
counterpropagation, the entangled photons are sent in different directions through
different atmospheric channels. The photons  reach the receivers
$\textrm{A}$ and $\textrm{B}$, where they  are analyzed with the corresponding polarization analyzers.

If the detector $\textrm{D}_{T_\mathrm{{A(B)}}}$ in the corresponding transmission 
channel clicks, this means that the incoming photon is linearly polarized with the angle 
$\theta_\mathrm{A(B)}$. If the detector $\textrm{D}_{R_\mathrm{{A(B)}}}$ in the 
corresponding reflection channel clicks, the incoming photon is assumed to be polarized 
in the orthogonal direction. For our purposes it is important to count the simultaneous 
clicks on the sites $\textrm{A}$ and $\textrm{B}$ in order to reconstruct the frequencies 
$P_{i_\mathrm{A},
i_\mathrm{B}}\!\left(\theta_\mathrm{A},\theta_\mathrm{B}\right)$, where $i_\mathrm{A(B)}{=}\{{T_\mathrm{A(B)}},{R_\mathrm{A(B)}}\}$. The frequency
	\begin{equation}
	P_\mathrm{same}\!\left(\theta_\mathrm{A},
	\theta_\mathrm{B}\right){=}P_{T_\mathrm{A},
		T_\mathrm{B}}\!\left(\theta_\mathrm{A},
	\theta_\mathrm{B}\right)+P_{R_\mathrm{A},
		R_\mathrm{B}}\!\left(\theta_\mathrm{A},
\theta_\mathrm{B}\right),\label{Eq:Psame}
	\end{equation}
of  simultaneous clicks of the pair of detectors
$\textrm{D}_{T_\mathrm{A}}$ and
$\textrm{D}_{T_\mathrm{B}}$  or
the pair $\textrm{D}_{R_\mathrm{A}}$ and $\textrm{D}_{R_\mathrm{B}}$,  and the frequency,	
	\begin{equation}
	P_\mathrm{different}\!\left(\theta_\mathrm{A},
	\theta_\mathrm{B}\right){=}P_{T_\mathrm{A},
		R_\mathrm{B}}\!\left(\theta_\mathrm{A},
	\theta_\mathrm{B}\right)+P_{R_\mathrm{A},
		T_\mathrm{B}}\!\left(\theta_\mathrm{A}, \theta_\mathrm{B}\right),
		\label{Eq:Pdifferent}
	\end{equation}
of simultaneous clicks of the pair of detectors
$\textrm{D}_{T_\mathrm{A}}$ and $\textrm{D}_{R_\mathrm{B}}$ or
$\textrm{D}_{R_\mathrm{A}}$ and $\textrm{D}_{T_\mathrm{B}}$ is then used for evaluation of the
correlation coefficients
\begin{equation}
E\left(\theta_\mathrm{A}, \theta_\mathrm{B}\right) =
\frac{P_\mathrm{same}\left(\theta_\mathrm{A},
	\theta_\mathrm{B}\right)-P_\mathrm{different}\left(\theta_\mathrm{A},
	\theta_\mathrm{B}\right)}{P_\mathrm{same}\left(\theta_\mathrm{A},
	\theta_\mathrm{B}\right)+P_\mathrm{different}\left(\theta_\mathrm{A},
	\theta_\mathrm{B}\right)}.\label{correlation}
\end{equation}

The Bell theorem in the Clauser-Horne-Shimony-Holt (CHSH)
form~\cite{BellReview,CHSH} states that for two sets of polarization angles the Bell
parameter
\begin{align}
\mathcal{B}&=\left|E\left(\theta_\mathrm{A}^{(1)},
\theta_\mathrm{B}^{(1)}\right)-E\left(\theta_\mathrm{A}^{(1)},
\theta_\mathrm{B}^{(2)}\right)\right|\label{BellParameter}
\\&+\left|E\left(\theta_\mathrm{A}^{(2)},
\theta_\mathrm{B}^{(2)}\right)+E\left(\theta_\mathrm{A}^{(2)},
\theta_\mathrm{B}^{(1)}\right)\right|\nonumber,
\end{align}
satisfies the condition
\begin{align}
	\mathcal{B}\leq 2,\label{BellIneq}
\end{align}
for local realistic theories. For nonclassical light fields the Bell inequality can be
violated. For such nonclassical fields the correlation properties are incompatible
with those of any local model. Violation of inequality (\ref{BellIneq})
certifies the presence of  nonlocal quantum correlations of light.

\subsection{Theoretical analysis}

An analysis of the Bell-inequality test requires derivation of the
joint probabilities, $P_{i_\mathrm{A}, i_\mathrm{B}}\!\left(\theta_\mathrm{A},\theta_\mathrm{B}\right)$,
for detection of photons at receiver sites $\textrm{A}$ and $\textrm{B}$. The 
derivation of these quantities can be performed on a similar footing as done in
Ref.~\cite{Semenov2010}. However, here we should not omit double clicks,
i.e., events where detectors in transmission and reflection channels click  
simultaneously at at least one site, $\textrm{A}$ or $\textrm{B}$. This may happen due to 
the stray light, dark counts and multi-photon pairs. According to the
photodetection theory~\cite{Mandel, Kelley} we can write the joint  probabilities of photon
detection as
\begin{align}
P_{i_\mathrm{A}, i_\mathrm{B}}\left(\theta_\mathrm{A},
\theta_\mathrm{B}\right)&=\Tr\left(
\hat{\Pi}_{i_\mathrm{A}}^{(c)}\hat{\Pi}_{i_\mathrm{B}}^{(c)}
\hat{\Pi}_{j_\mathrm{A}}^{(0)}\hat{\Pi}_{j_\mathrm{B}}^{(0)}
\hat\rho\right)\nonumber\\
&+\frac{1}{2}\Tr\left(
\hat{\Pi}_{i_\mathrm{A}}^{(c)}\hat{\Pi}_{i_\mathrm{B}}^{(c)}
\hat{\Pi}_{j_\mathrm{A}}^{(c)}\hat{\Pi}_{j_\mathrm{B}}^{(0)}
\hat\rho\right)\nonumber\\
&+\frac{1}{2}\Tr\left(
\hat{\Pi}_{i_\mathrm{A}}^{(c)}\hat{\Pi}_{i_\mathrm{B}}^{(c)}
\hat{\Pi}_{j_\mathrm{A}}^{(0)}\hat{\Pi}_{j_\mathrm{B}}^{(c)}
\hat\rho\right)\nonumber\\
&+\frac{1}{4}\Tr\left(
\hat{\Pi}_{i_\mathrm{A}}^{(c)}\hat{\Pi}_{i_\mathrm{B}}^{(c)}
\hat{\Pi}_{j_\mathrm{A}}^{(c)}\hat{\Pi}_{j_\mathrm{B}}^{(c)}\hat\rho\right),
\label{ProbabilitySquash}
\end{align}
where $\hat{\rho}$ is the density operator and
\begin{align}
&\hat{\Pi}_{i_\mathrm{A(B)}}^{(0)}=:\exp\left(-\eta_c\,
\hat{a}^\dagger_{i_\mathrm{A(B)}}\hat{a}_{i_\mathrm{A(B)}}-\nu
\right):,\label{POVM}\\
&\hat{\Pi}_{i_\mathrm{A(B)}}^{(c)}=1-:\exp\left(-\eta_c\,
\hat{a}^\dagger_{i_\mathrm{A(B)}}\hat{a}_{i_\mathrm{A(B)}}-\nu\right):\label{POVM1}
\end{align}
are the positive operator-valued measures for the on/off detectors $i_\mathrm{A(B)}$,
 related to the absence and presence of detection events,
respectively.
Here
$\hat{a}_{i_\mathrm{A(B)}}$ and $\hat{a}^\dagger_{i_\mathrm{A(B)}}$ are photonic
annihilation and
creation operators, respectively, for the field modes at the
$i_\mathrm{A(B)}{=}\{T_\mathrm{A(B)},R_\mathrm{A(B)}\}$ output of the
$\textrm{PBS}$ such that $i_\mathrm{A(B)}{\neq} j_\mathrm{A(B)}$, $\eta_c$ is the
detection efficiency, $\nu$ is the mean number of stray-light
and dark counts~\cite{Semenov2008}, and $:...:$ denotes the normal ordering. 

The last three terms in Eq.~(\ref{ProbabilitySquash}) describe the contributions from
double-click events, when both detectors on at least one side click. For
such events we assign random values of the corresponding qubits. In practice this 
means that if both detectors, $\textrm{D}_{T_\mathrm{{A(B)}}}$ and 
$\textrm{D}_{R_\mathrm{{A(B)}}}$, at site $\textrm{A(B)}$ click simultaneously, we 
randomly ascribe to such a realization the event related to the click on detector 
$\textrm{D}_{T_\mathrm{{A(B)}}}$ or $\textrm{D}_{R_\mathrm{{A(B)}}}$ with probability $1/2$. In the case where all four detectors click 
simultaneously we randomly ascribe to this realization any of four possible pair events with  
probability $1/4$.  The three terms associated with double-click events have not been 
considered in Ref.~\cite{Semenov2010}. In the present article we analyze the Bell inequality taking these contributions into account.

We now specify the quantum state, 
$\hat{\rho}{=}\ket{\mathrm{PDC}}\bra{\mathrm{PDC}}$, generated  by the PDC source
(cf.~Refs.~\cite{Ma,Kok}):
\begin{equation}
\ket{\mathrm{PDC}}=(\cosh\xi)^{-2}\sum\limits_{n=0}^{+\infty}
\sqrt{n+1}\tanh^n\xi\left|\Phi_n\right\rangle.\label{PDC1}
\end{equation}
Herein $\xi$ is the squeezing parameter,
\begin{align}
&\left|\Phi_n\right\rangle=\label{PDC2}\\
&\frac{1}{\sqrt{n+1}}\sum\limits_{m=0}^{n}(-1)^m
\left|n{-}m\right\rangle_\mathrm{H_A}\left|m\right\rangle_\mathrm{V_A}
\left|m\right\rangle_\mathrm{H_B}\left|n{-}m\right\rangle_\mathrm{V_B},\nonumber
\end{align}
and $\left|n\right\rangle_\mathrm{H_{A(B)}}$ and $\left|n\right\rangle_\mathrm{V_{A(B)}}$ are the
 photon-number states of the horizontal and vertical polarization modes, respectively, sent 
 to
the receiver
$\textrm{A(B)}$. For small values of $\xi$ the first two terms in
Eq.~(\ref{PDC1}) are the most relevant, such that $\left|\Phi_0\right\rangle$ is the vacuum state and
\begin{align}
&\ket{\Phi_1}=\label{BellState1}\\&\frac{1}{\sqrt{2}}\Big(\ket{1}_\mathrm{H_A}\ket{0}_\mathrm{V_A}
\ket{0}_\mathrm{H_B}\ket{1}_\mathrm{V_B}-\ket{0}_\mathrm{H_A}\ket{1}_\mathrm{V_A}
\ket{1}_\mathrm{H_B}\ket{0}_\mathrm{V_B}\Big)\nonumber
\end{align}
is the  Bell state,
\begin{align}
\ket{\mathcal{B}}\equiv\ket{\Phi_1}=\frac{1}{\sqrt{2}}\Big(\ket{\mathrm{H}}_\mathrm{A}
\ket{\mathrm{V}}_\mathrm{B}-
\ket{\mathrm{V}}_\mathrm{A}\ket{\mathrm{H}}_\mathrm{B}
\Big),\label{BellState2}
\end{align}
which maximally violates inequality~(\ref{BellIneq}).
Here the state
$\ket{\mathrm{H}}_\mathrm{A(B)}=\ket{1}_\mathrm{H_{A(B)}}\ket{0}_\mathrm{V_{A(B)}}$
corresponds to the photon in the horizontal mode at site A(B), whereas the state
$\ket{\mathrm{V}}_\mathrm{A(B)}=\ket{0}_\mathrm{H_{A(B)}}\ket{1}_\mathrm{V_{A(B)}}$
corresponds to the photon in the vertical mode at site A(B).

Further analysis is performed similarly to that in Ref.~\cite{Semenov2010};
for details of calculations see also Appendixes~\ref{App:Coefficients} and
\ref{App:Bell states}. First, we should modify the initial quantum state [cf.~Eqs.~(\ref{PDC1}) or
(\ref{BellState1})] according to the quantum-state
input-output relation for fading channels (cf.~Ref.~\cite{Semenov2009,Vasylyev2012})
with fluctuating transmission efficiencies (transmittances) $\eta_\mathrm{A}$ for 
modes $\textrm{H}_A$ $\textrm{V}_A$ and $\eta_\mathrm{B}$ for modes
$\textrm{H}_B$ $\textrm{V}_B$. This relation has a simple form in the Glauber-Sudarshan
$P$ representation,
\begin{equation}
P_\mathrm{out}\left(\boldsymbol{\alpha}\right)=
\int\limits_{0}^{1}\D^2
\boldsymbol{\eta}\frac{1}{\eta_\mathrm{A}^2\eta_\mathrm{B}^2}
\mathcal{P}\left(\eta_\mathrm{A},\eta_\mathrm{B}\right)
P_\mathrm{in}\left(\boldsymbol{{\alpha}_{\eta}}\right),
\end{equation}
where $P_\mathrm{in}$ and $P_\mathrm{out}$ are the Glauber-Sudarshan $P$ functions 
(cf.~Ref.~\cite{Glauber,GlauberPRA,Sudarshan})
of light at the source [corresponding to the state~(\ref{PDC1}) or 
(\ref{BellState1})] and at the receivers, respectively,
$\boldsymbol{\alpha}=\left(\alpha_{\scriptscriptstyle H_\mathrm{A}},
\alpha_{\scriptscriptstyle V_\mathrm{A}},
\alpha_{\scriptscriptstyle H_\mathrm{B}},\alpha_{\scriptscriptstyle V_\mathrm{B}}\right)$ and
$\boldsymbol{{\alpha}_{\eta}}=\left(\alpha_{\scriptscriptstyle 
H_\mathrm{A}}/\sqrt{\eta}_\mathrm{A},
\alpha_{\scriptscriptstyle V_\mathrm{A}}/\sqrt{\eta}_\mathrm{A},
\alpha_{\scriptscriptstyle H_\mathrm{B}}/\sqrt{\eta}_\mathrm{B},\alpha_{\scriptscriptstyle 
V_\mathrm{B}}/\sqrt{\eta}_\mathrm{B}\right)$,
 $\D^2
\boldsymbol{\eta}=\D \eta_\mathrm{A}\D \eta_\mathrm{B}$. In this equation the probability
distribution of the transmittance (PDT),
$\mathcal{P}\left(\eta_\mathrm{A},\eta_\mathrm{B}\right)$, is the main characteristics of
the atmospheric channels.

The explicit form of the PDT depends on the characteristics of the irradiated light beam,
radius of the receiver aperture, and characteristics 
of the atmospheric channel, i.e., its length and turbulence
conditions. For characterization of channels with homogeneous and isotropic turbulence we use the Rytov parameter 
(cf.~Refs.~\cite{Tatarskii,Ishimaru,Andrews,Andrews2}),
\begin{align}
 \sigma_{R}^2=1.23 C_n^2 k^{\frac{7}{6}}L^{\frac{11}{6}},\label{RytovParameter}
\end{align}
which quantifies the integral effect of optical turbulence on the whole 
channel.
Here $C_n^2$ is the index-of-refraction structure constant, characterizing the local 
strength of turbulence, $k$ is the wave number of 
optical radiation, and $L$ is the channel length. Throughout this article, we 
consider weak- to moderate-turbulence channels ($\sigma_{ 
R}^2{\approx}1...10$) and strong-turbulence
($\sigma_{ R}^2{\gg}1$) channels. These notions characterize the integral effects of the turbulence on the quantum light over the whole propagation paths under study.

Derivation of the PDT applies the knowledge of classical atmospheric optics (see e.g.
\cite{Tatarskii, Ishimaru,Andrews,Andrews2,Fante1,Fante2,Chumak2006,Chumak2016}). 
For weak-turbulence channels, $\sigma_{R}{<}1$,
when the leading disturbance is beam wandering, the PDT takes the form of
the log-negative Weibull distribution (see~Ref.~\cite{Vasylyev2012} for its derivation). For weak- to moderate- and strong-turbulence channels, deformations of
the beam play an important role. The corresponding PDT in the elliptic-beam 
approximation has been derived in Ref.~\cite{Vasylyev2016}. It is important to note that 
under strong-turbulence conditions the elliptic-beam model is in reasonable agreement 
with the log-normal distribution~\cite{Diament, Perina,Perina1973,Milonni}, provided that 
the latter is restricted
to the	physical domain, $\eta_\mathrm{A(B)}{\leq}1$. Such behavior has been
experimentally verified in Ref.~\cite{Capraro}.  In the present paper we apply two scenarios
with the corresponding PDTs. First, we consider a 1.6\,km channel in the city of Erlangen
(cf.~Ref.~\cite{Usenko}) as a typical example of weak- to moderate-turbulence channels. The corresponding
experimental results are in good agreement with the recently proposed elliptic-beam 
model for the PDT
(see~Ref.~\cite{Vasylyev2016} and Appendix~\ref{App:WeakChannel} for more details). Second, we consider a
144-km channel on the Canary Islands (cf.~\cite{Capraro}) as a typical strong-turbulence channel. 
For simplicity, the corresponding PDT is approximated in the following by the truncated log-normal 
distribution (for more details see Appendix~\ref{App:StrongChannel}).

For the purposes 
of this paper it is important to note that within the given apertures the weak- to moderate-turbulence channel is characterized by a small value of the fluctuation of losses, 
$\langle\Delta\eta^2\rangle/\langle\eta\rangle^2$. The strong-turbulence channel has a 
large value of this parameter. It is noteworthy that higher moments of the transmittance 
may also play a crucial role for characterization of the considered channels.

As the next step, we  rewrite the state in terms of transmitted,
$T_{\mathrm A(B)}$, and reflected, $R_{\mathrm A(B)}$, modes in the inputs of
polarization-analyzer detectors.
The corresponding input-output relations for the field operators read as
\begin{align}
&\hat{a}_\mathrm{\scriptscriptstyle
	H_{A(B)}}=\hat{a}_{\scriptscriptstyle{T_\mathrm{A(B)}}}\cos\theta_\mathrm{A(B)}-
\hat{a}_{\scriptscriptstyle{R_\mathrm{A(B)}}}
\sin\theta_\mathrm{A(B)}\label{IORop1},\\
&\hat{a}_{\scriptscriptstyle\mathrm{V_{A(B)}}}=
\hat{a}_{\scriptscriptstyle{T_\mathrm{A(B)}}}\sin\theta_\mathrm{A(B)}+
\hat{a}_{\scriptscriptstyle{R_\mathrm{A(B)}}}\cos\theta_\mathrm{A(B)}\label{IORop2},
\end{align}
where $\hat{a}_{\scriptscriptstyle\mathrm{H_{A(B)}}}$ and $\hat{a}_{\scriptscriptstyle
	\mathrm{V_{A(B)}}}$ are field annihilation operators for the site $\textrm{A(B)}$
	horizontal and vertical modes, respectively. Technical details of applications of these
	relations can be found in Ref.~\cite{Semenov2010} as well as in
	Appendixes~\ref{App:Coefficients}, and \ref{App:Bell states}.
The obtained density operator is substituted in
Eq.~(\ref{ProbabilitySquash}) and then in Eqs.~(\ref{Eq:Psame}) and
(\ref{Eq:Pdifferent}), which leads to explicit forms of
$P_\mathrm{same}\left(\theta_\mathrm{A},\theta_\mathrm{B}\right)$
and $P_\mathrm{different}\left(\theta_\mathrm{A},\theta_\mathrm{B}\right)$.  Finally, we
use these probabilities for calculations of the correlation
coefficients, (\ref{correlation}), and maximization of the Bell
parameter, (\ref{BellParameter}), with respect to the angles $\theta_\mathrm{A}$  and
$\theta_\mathrm{B}$.

\subsection{Parametric down-conversion source}

In the case of the PDC source the initial state is given by the
density operator $\hat{\rho}=\ket{\mathrm{PDC}}\bra{\mathrm{PDC}}$
[cf.~Eq.~(\ref{PDC1})]. Applying the above discussed analysis (cf. also
Appendix~\ref{App:Coefficients}), one gets for the probabilities
$P_\mathrm{same}\left(\theta_\mathrm{A},\theta_\mathrm{B}\right)$
and $P_\mathrm{different}\left(\theta_\mathrm{A},\theta_\mathrm{B}\right)$
\begin{widetext}
\begin{align}
&P_\mathrm{i}\left(\theta_\mathrm{A},
\theta_\mathrm{B}\right)=\frac{1}{2}+\frac{e^{-4\nu}}{2}
\left(1-\tanh^2\xi\right)^4\label{ProbabilitySpecialSame2}\\
&{\times}\left[e^{2\nu}\left(2\left\langle
\frac{1}{C_\mathrm{0}+C_\mathrm{1A}+C_\mathrm{1B
}+C_\mathrm{i}}\right\rangle
{-}\left\langle\frac{C_\mathrm{0}}{{(C_\mathrm{0}+C_\mathrm{1A})}^2}\right\rangle{-}
\left\langle\frac{C_\mathrm{0}}{{(C_\mathrm{0}+C_\mathrm{1B})}^2}\right\rangle{-}
2\left\langle\frac{1}{C_\mathrm{0}+C_\mathrm{1A}+C_\mathrm{1B}+C_\mathrm{
j}}\right\rangle\right){+}\left\langle\frac{1}{C_\mathrm{0}}\right\rangle\right]
.\nonumber
\end{align}
Here $i,j{=}\{\mathrm{same},\mathrm{different}\}$, $i{\neq}j$,
\begin{equation}
C_\mathrm{0}=\left\{\eta_c^2\eta_\mathrm{A}\eta_\mathrm{B}\tanh^2\xi-
\left[1+\left(\eta_c \eta_\mathrm{A}-1\right)\tanh^2\xi\right]\left[1+\left(\eta_c
\eta_\mathrm{B}-1\right)\tanh^2\xi\right]\right\}^2,\label{Eq:C02}
\end{equation}
\begin{eqnarray}
C_\mathrm{1A(B)}=&\eta_c \eta_\mathrm{B(A)}&\left(1-\eta_c
\eta_\mathrm{A(B)}\right)\left(1-\tanh^2\xi\right)\tanh^2\xi\label{Eq:C1A}
\\
&\times&\left\{\eta_c^2\eta_\mathrm{A(B)}\eta_\mathrm{B(A)}\tanh^2\xi-
\left[1+\left(\eta_c
\eta_\mathrm{A(B)}-1\right)\tanh^2\xi\right]\left[1+\left(\eta_c
\eta_\mathrm{B(A)}-1\right)\tanh^2\xi\right]\right\},\nonumber
\end{eqnarray}
\begin{eqnarray}
C_\mathrm{same}=\eta_c^2
\eta_\mathrm{A} \eta_\mathrm{B}\tanh^2\xi\left(1-\tanh^2\xi\right)
^2\label{Eq:Csame2}
\left[\left(1-\eta_c \eta_\mathrm{A}\right)\left(1-\eta_c \eta_\mathrm{B}\right)
\tanh^2\xi-\sin^2
\left(\theta_\mathrm{A}-\theta_\mathrm{B}\right)\right],
\end{eqnarray}
\begin{eqnarray}
C_\mathrm{different}=\eta_c^2
\eta_\mathrm{A} \eta_\mathrm{B}\tanh^2\xi\left(1-\tanh^2\xi\right)
^2\label{Eq:Cdifferent2}
\left[\left(1-\eta_c \eta_\mathrm{A}\right)\left(1-\eta_c \eta_\mathrm{B}\right)
\tanh^2\xi-\cos^2
\left(\theta_\mathrm{A}-\theta_\mathrm{B}\right)\right],
\end{eqnarray}
\end{widetext}
and 
\begin{align}
 \left\langle\ldots\right\rangle=\int\limits_{0}^{1}\D^2
\boldsymbol{\eta}\ldots
\mathcal{P}\left(\eta_\mathrm{A},\eta_\mathrm{B}\right)\label{Eq:mean}
\end{align}
means averaging by the channel transmittances.

\subsection{Bell-state source}

   For the case of a weak-intensity source, the state at the transmitter can be effectively
   approximated by the Bell states [cf.~Eqs.~(\ref{BellState1}) and (\ref{BellState2})]. In this
   case calculation of the Bell parameter can be performed explicitly. The maximal
   value of the Bell parameter [cf.~Eq.~(\ref{BellParameter})] reads as
\begin{align}
&\mathcal{B}=\label{S_OnOff}\\&\frac{2\sqrt{2}\,p_{\scriptscriptstyle
\mathcal{B}}\eta_c^2e^{2\nu}}
{p_{\scriptscriptstyle
\mathcal{B}}\left(e^{2\nu}{+}\eta_c{-}1\right)^2{+}p_{\scriptstyle\mathrm{0}}
\left(e^{2\nu}{-}1\right)^2{+}
p_{\scriptstyle 1}\left(e^{2\nu}{-}1\right)
\left(e^{2\nu}{+}\eta_c{-}1\right)}.\nonumber
\end{align}
Here
\begin{equation}
p_{\scriptscriptstyle \mathcal{B}}=\left\langle
\eta_\mathrm{A} \eta_\mathrm{B}\right\rangle\label{Eq:pB}
\end{equation}
is the probability that at the output one  gets the Bell state,
\begin{equation}
p_{\scriptstyle\mathrm{0}}=\left\langle
\left(1-\eta_\mathrm{A}\right)
\left(1-\eta_\mathrm{B}\right)\right\rangle\label{Eq:p0}
\end{equation}
is the probability that both photons do not reach the receiver, and finally
\begin{equation}
p_{\scriptstyle 1}=\left\langle \eta_\mathrm{A}
\left(1-\eta_\mathrm{B}\right)\right\rangle+
\left\langle \eta_\mathrm{B}
\left(1-\eta_\mathrm{A}\right)\right\rangle\label{Eq:p1}
\end{equation}
is the probability that only one photon will reach the receiver.
Details of calculations can be found in Appendix~\ref{App:Bell states}.

\section{Copropagation}
\label{Copropagation}

The scenario of copropagation, which is represented in Fig.~\ref{Scheme}(a) and was
experimentally studied in \cite{Fedrizzi},  is characterized by a very short time interval
between two entangled pulses. This interval is much less than the time, for which the
atmosphere is changed.  As a result, the two transmittances $\eta_\mathrm{A}$
and $\eta_\mathrm{B}$ can be considered to be completely correlated. The
PDT in this case is given by
\begin{align}
	\mathcal{P}\left(\eta_\mathrm{A},\eta_\mathrm{B}\right)=
	\mathcal{P}\left(\eta_\mathrm{A}\right)\delta\left(\eta_\mathrm{A}-\eta_\mathrm{B}\right),
	\label{PDT_Copropagation}
\end{align}
where $\mathcal{P}\left(\eta_\mathrm{A}\right)$ is the single-mode PDT. This
implies that Eq.~(\ref{Eq:mean}) can be rewritten as
\begin{align}
 \left\langle\ldots\right\rangle=\int\limits_{0}^{1}\D
\eta\ldots
\mathcal{P}\left(\eta\right),\label{Eq:meanco}
\end{align}
where we  explicitly take that
$\eta_\mathrm{A}{=}\eta{=}\eta_\mathrm{B}$.

As mentioned in Sec.~\ref{Introduction}, such fading channels
demonstrate higher values of the Bell parameter compared with deterministic
attenuation channels characterized by the transmittance
$\eta_0{=}\langle \eta\rangle$
(cf.~Ref.~\cite{Semenov2010}). Indeed, from the Cauchy-Schwarz
inequality it follows that $\left\langle \eta^2 \right\rangle{\geq}\left\langle
\eta\right\rangle^2$. This yields that the probability of
preserving Bell states by the channel [cf.~Eq.~(\ref{Eq:pB})] is higher for the
correlated fading channel compared with the deterministic-loss channel, i.e.,
\begin{equation}
p_{\scriptscriptstyle \mathcal{B}}=\left\langle \eta^2 \right\rangle \geq
\eta_0^2=
p_{\scriptscriptstyle
\mathcal{B}}^\mathrm{det}.\label{Eq:IneqCoprop}
\end{equation}
As a result, the violation of the Bell inequalities is more significant in the case of
correlated atmospheric channels.

\begin{figure}[ht!]
\includegraphics[clip=,width=\linewidth]{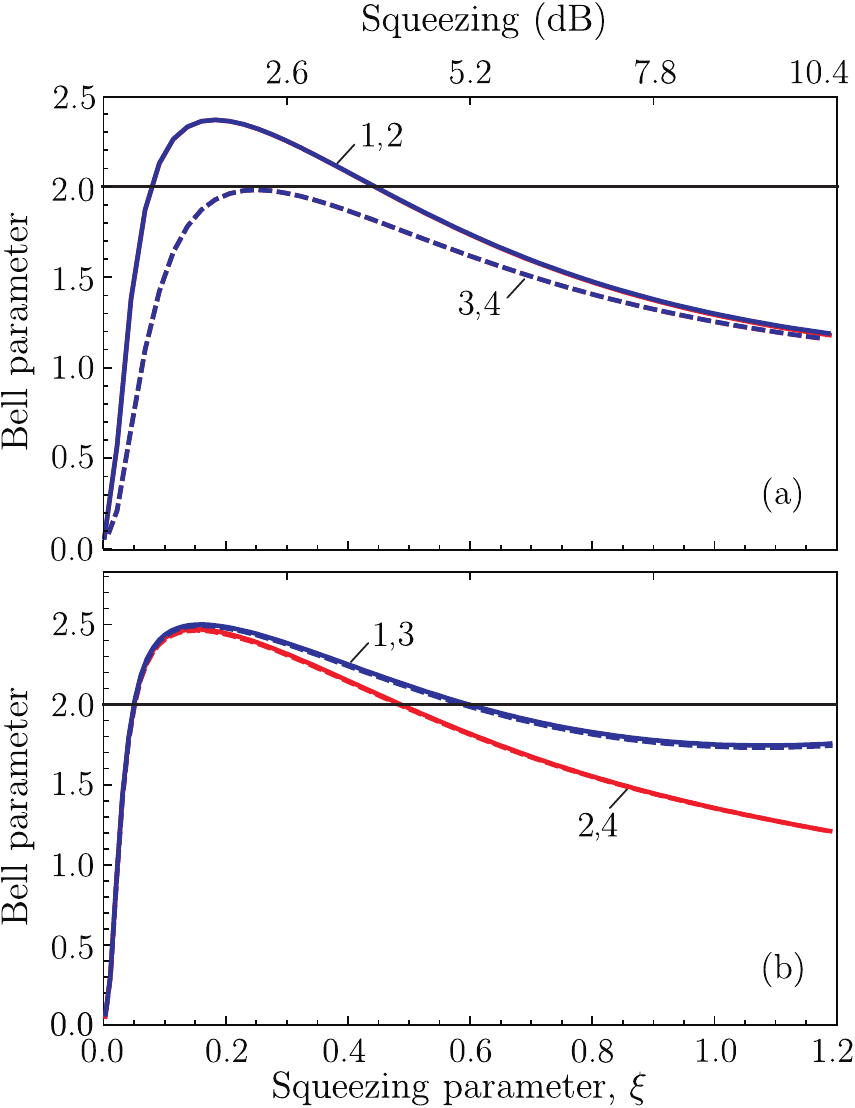}
\caption{\label{Fig2} (Color online) The Bell parameter, $\mathcal{B}$, vs the
squeezing parameter, $\xi$, for the scenario of copropagation. The PDT is taken
(a) for the strong-turbulence channel (cf.~Appendix~\ref{App:StrongChannel}), where
the mean number of stray-light and dark counts is 
$\nu{=}1.7{\times}10^{{-}5}$, and (b) for the weak- to moderate-turbulence channel 
(cf.~Appendix~\ref{App:WeakChannel}), with $\nu{=}3{\times}10^{{-}3}$. 
For both cases, lines 1 and 2 correspond to fading channels without and with the incorporation 
of double-clicks events, respectively. Lines 3 and 4 correspond to the 
deterministic-loss channels with $\eta_0=\langle\eta\rangle$, without and with the 
incorporation of double-clicks events, respectively. The detection efficiency is 
$\eta_c=0.3$, which includes 3-dB losses at the beam splitter of the receiver. Here 
and in the following figures, the horizontal line $\mathcal{B}=2$ shows the local-realism bound.}
\end{figure}

In Ref.~\cite{Semenov2010} this property has been theoretically considered by neglecting
the double-click events, i.e., Eq.~(\ref{ProbabilitySquash}) has been taken without the last three terms (see
also Appendix~\ref{App:NDubleClicks}). However, the incorporation of double-click events
may diminish the measured value of the Bell parameter. In the following we consider these effects.

In Fig.~\ref{Fig2} we demonstrate the dependence of the Bell parameter, 
$\mathcal{B}$, on the squeezing parameter, $\xi$,  of the PDC 
state [cf.~Eq.~(\ref{PDC1})] for the case of incorporation of double-click events and for 
discarding it (see~Ref.~\cite{Semenov2010} and Appendix~\ref{App:NDubleClicks} for the 
latter). An interesting result is that in the case of large atmospheric losses, such as occur in 
the considered strong-turbulence channel (cf.~Appendix~\ref{App:StrongChannel}),
the effect of double-click events is negligible. This can be explained by the relatively low probability of multiphoton pairs' passing through the high-loss channel even for a bright source (large squeezing parameter, $\xi$).
A similar behavior is observed for a weak-intensity source (small squeezing parameter, $\xi$) in the case of 
weak- to moderate-turbulence channel (cf.~Appendix~\ref{App:WeakChannel}).
In the latter case the overall losses are relatively small. As a result, multiphoton pairs 
are passed through the channel with a higher probability.  For this scenario, they 
significantly contribute to the double-click events. We ascribe to each such event a 
random value of the corresponding qubit. This protocol is characterized by classical 
probabilities and it certainly diminishes the nonclassical correlations between site 
$\mathrm{A}$ and site $\mathrm{B}$.  Consequently, the value of the Bell parameter decreases 
more rapidly with increasing squeezing parameter $\xi$ compared to the case of discarding 
the double-click events.

Another important issue that appears in the scenario of copropagation, is the fact that for weak-to-moderate turbulence channels the advantages of the atmospheric links are negligible. For weak-intensity sources, when the state is approximately equal to the Bell state~(\ref{BellState2}), this fact can be easily explained by the low ratio $\langle\Delta\eta^2\rangle/\langle\eta\rangle^2$. Indeed, in this case the equality in Eq.~(\ref{Eq:IneqCoprop}) is satisfied almost exactly. However, as we can see in Fig.~\ref{Fig2}(b), even the presence of multiphoton pairs does not improve the value of the Bell parameter.

\section{Counterpropagation}
\label{Counterpropagation}

The scenario of counterpropagation, which is represented in
Fig.~\ref{Scheme}(b), corresponds to channels with completely uncorrelated
transmittances, $\eta_\mathrm{A}$ and $\eta_\mathrm{B}$. The PDT in this
case reads as
\begin{align}
	\mathcal{P}\left(\eta_\mathrm{A},\eta_\mathrm{B}\right)=
\mathcal{P}_\mathrm{A}\!\left(\eta_\mathrm{A}\right)\mathcal{P}_\mathrm{B}\!
\left(\eta_\mathrm{B}\right),
	\label{PDT_Counterpropagation}
\end{align}
where
$\mathcal{P}_\mathrm{A}\!\left(\eta_\mathrm{A}\right)$ and
$\mathcal{P}_\mathrm{B}\!\left(\eta_\mathrm{B}\right)$ are the single-mode PDT for
the corresponding channels. As a particular case, this scenario describes the
situation when only one mode is sent through a turbulent atmosphere while
another one is operated near the source, such as experimentally studied
in Ref.~\cite{Ursin}. In this case
\begin{align}
 \mathcal{P}_\mathrm{B}\!
\left(\eta_\mathrm{B}\right)=\delta\left(\eta_\mathrm{B}-\eta_\mathrm{b}\right),
\end{align}
where $\eta_\mathrm{b}$ is the deterministic transmission coefficient of 
channel $\textrm{B}$. 

In the case of a weak-intensity source, which is effectively described by the Bell
state~(\ref{BellState1}), Eqs.~(\ref{Eq:p0}), (\ref{Eq:p1}), and (\ref{Eq:pB}) yield
 \begin{align}
& p_{\scriptscriptstyle\mathrm{0}}=
 \left(1-\langle \eta_\mathrm{A}\rangle\right)
 \left(1-\langle \eta_\mathrm{B}\rangle\right),\label{Eq:p0unc}\\
& p_1=\langle \eta_\mathrm{A}\rangle
 \left(1-\langle \eta_\mathrm{B} \rangle\right)+
 \langle \eta_\mathrm{B}\rangle
 \left(1-\langle \eta_\mathrm{A}\rangle\right),\label{Eq:p1unc}\\
& p_{\scriptscriptstyle \mathcal{B}}=\langle
 \eta_\mathrm{A}\rangle\langle \eta_\mathrm{B}\rangle.\label{Eq:pBunc}
 \end{align}
This means that the corresponding fading channel is completely equivalent to the
deterministic-loss channel with transmittances $\eta_\mathrm{a(b)}{=}\langle
\eta_\mathrm{A(B)} \rangle$.  Hence, in this case fading does not result in any advantages.

\begin{figure}[ht!]
\includegraphics[clip=,width=\linewidth]{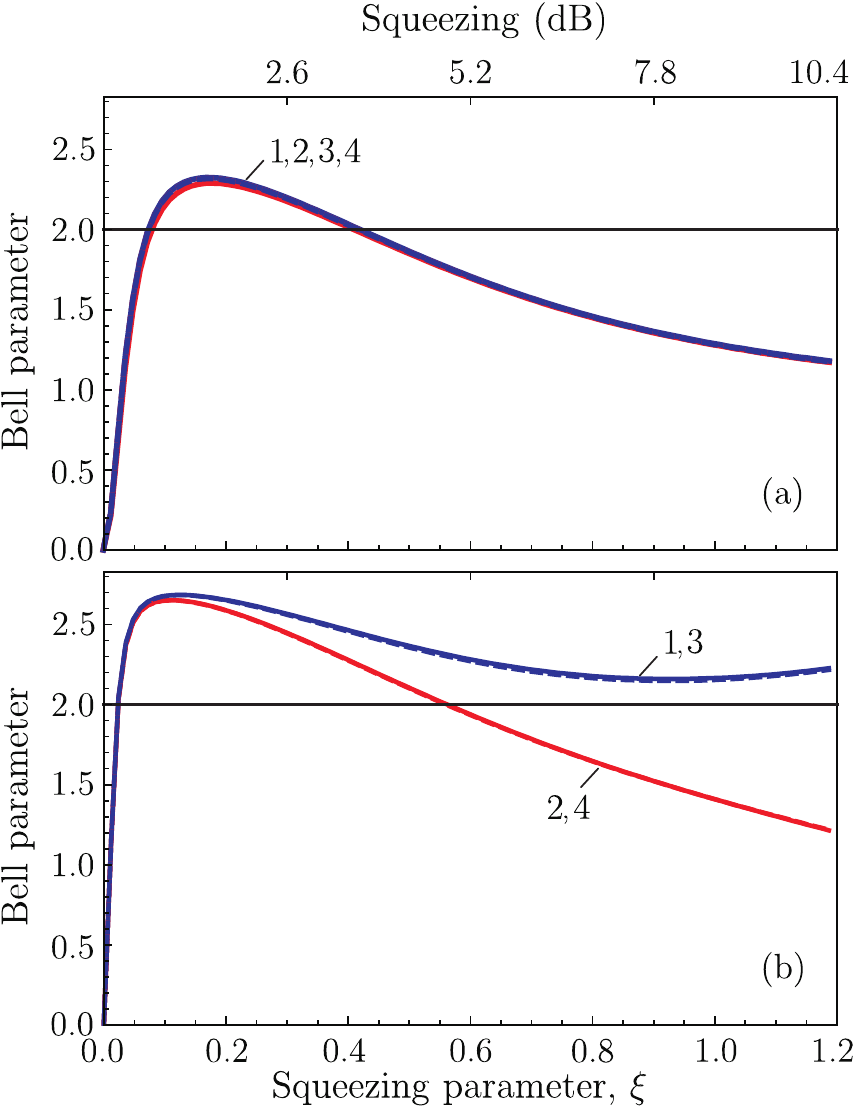}
\caption{\label{Fig3}  (Color online)  The Bell parameter, $\mathcal{B}$, vs the
squeezing parameter, $\xi$, for the scenario of counterpropagation. The PDT is taken
(a) for the strong-turbulence channel (cf.~Appendix~\ref{App:StrongChannel}), where the 
mean number of stray-light and dark counts is $\nu{=}1.7{\times}10^{{-}5}$, and (b) for the weak- to moderate-turbulence channel (cf.~Appendix~\ref{App:WeakChannel}), with $\nu{=}3{\times}10^{{-}3}$. 
For both cases, lines 1 and 2 correspond to fading channels without and with the incorporation 
of double-clicks events, respectively. Lines 3 and 4 correspond to the 
deterministic-loss channels with 
$\eta_a=\eta_b=\langle\eta_\mathrm{A}\rangle=\langle\eta_\mathrm{B}\rangle$, without and with the incorporation of double-clicks events, respectively. The detection efficiency is $\eta_c{=}0.6$.}
\end{figure}

Our study has shown that the atmospheric links for the considered channels  give the same result as the related deterministic-loss channels even for strong-intensity sources, when contributions from multiphoton pairs are essential. Results of the corresponding calculations are
presented in Fig.~\ref{Fig3}. It is also important to note that the effect of double-click events is visible only for strong-intensity sources (for large values of the squeezing parameter $\xi$) with weak- to moderate-turbulence channels [cf.~Fig~\ref{Fig3}(b)]. In all other cases the amount of contributions of multi-photon pairs, stray light, and dark counts at the receiver is not enough to diminish the value of the Bell parameter essentially.

\section{Postselection procedure}
\label{Sec:Post-processing}

\begin{figure}[ht!]
	\includegraphics[clip=,width=\linewidth]{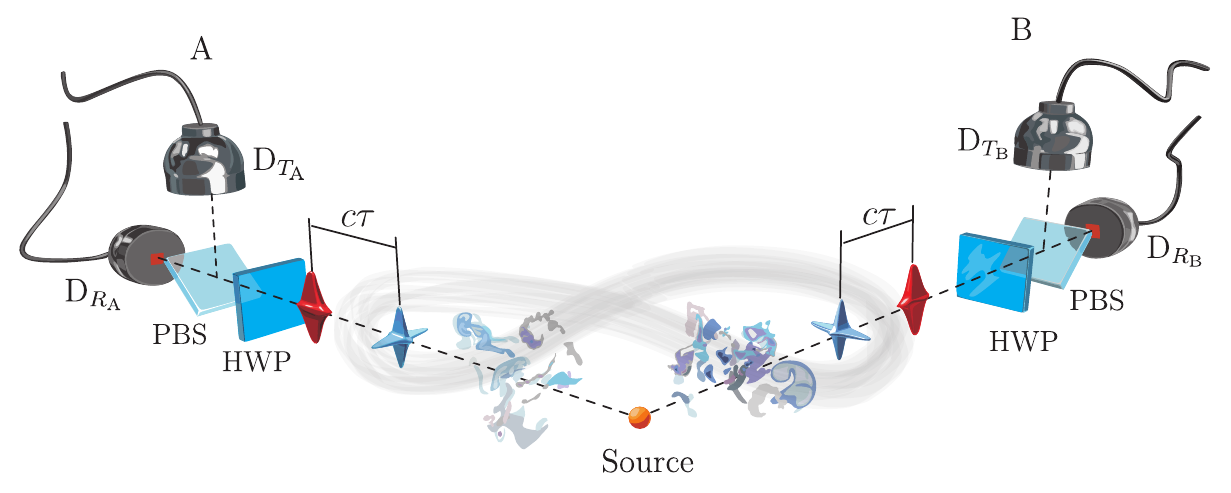}
	\caption{\label{Fig4}  (Color online) Bell-inequality test with the postselection procedure 
	proposed in Ref.~\cite{Capraro}. Strong light pulses are sent 
	in front of the nonclassical-light pulses in order to test the transmittance of the 
	channels. The time $\tau$ is much less than the typical time of atmospheric variations. 
	One considers only those events for which on both sides the measured transmittance 
	$\eta_\mathrm{A(B)}\geq\eta_\mathrm{ps}$, where $\eta_\mathrm{ps}$ is the 
	postselection threshold.}
\end{figure}

In cases where verifications of Bell-inequality violations are impossible, one can try to improve the 
situation with a certain postselection procedure, using the technique proposed in 
Ref.~\cite{Capraro}. For this purpose one sends intense light pulses in each channel, 
before the series of nonclassical-light pulses. With these pulses one can test the channel 
and then postselect the events with transmittances, 
$\eta_\mathrm{A(B)}\geq\eta_\mathrm{ps}$, exceeding a certain postselection threshold, 
$\eta_\mathrm{ps}$ (see Fig.~\ref{Fig4}). In this case, the time $\tau$ between the test and 
the nonclassical pulses should be smaller then the timescale of atmospheric variations.   

The theoretical analysis of this scheme is based on the corresponding reformulation of the PDT. The single-mode PDT after postselection, $\mathcal{\widetilde{P}}_\mathrm{A(B)}\!\left(\eta_\mathrm{A(B)}\right)$, is given by
\begin{align}
 \mathcal{\widetilde{P}}_\mathrm{A(B)}\!\left(\eta_\mathrm{A(B)}\right)=\frac{1}{\overline{F}_\mathrm{A(B)}\!\left(\eta_\mathrm{ps}\right)}\mathcal{P}_\mathrm{A(B)}\!\left(\eta_\mathrm{A(B)}\right)\label{PDT_PS}
\end{align}
for $\eta_\mathrm{A(B)}\in[\eta_\mathrm{ps},1]$ and 0 elsewhere. Here
\begin{align}
 \overline{F}_\mathrm{A(B)}\!\left(\eta_\mathrm{ps}\right)=\int\limits_{\eta_\mathrm{ps}}^{1}\D\eta\mathcal{P}_\mathrm{A(B)}\!\left(\eta\right)
\end{align}
is the exceedance (complementary cumulative probability distribution) of the PDT, which is the probability that the transmittance will exceed the value of $\eta_\mathrm{ps}$. The 
calculations for the postselection procedure are performed with the PDT 
given in Eq.~(\ref{PDT_PS}). It follows that 
\begin{align}
 \overline{F}_\mathrm{AB}\!\left(\eta_\mathrm{ps}\right)=\overline{F}_\mathrm{A}\!\left(\eta_\mathrm{ps}\right)
 \overline{F}_\mathrm{B}\!\left(\eta_\mathrm{ps}\right)\label{Eq:Exceed2}
\end{align}
is the probability that the transmittances exceed the value of $\eta_\mathrm{ps}$ in both channels. This quantity characterizes the feasibility of the postselection procedure.  

\begin{figure}[ht!]
\includegraphics[clip=,width=\linewidth]{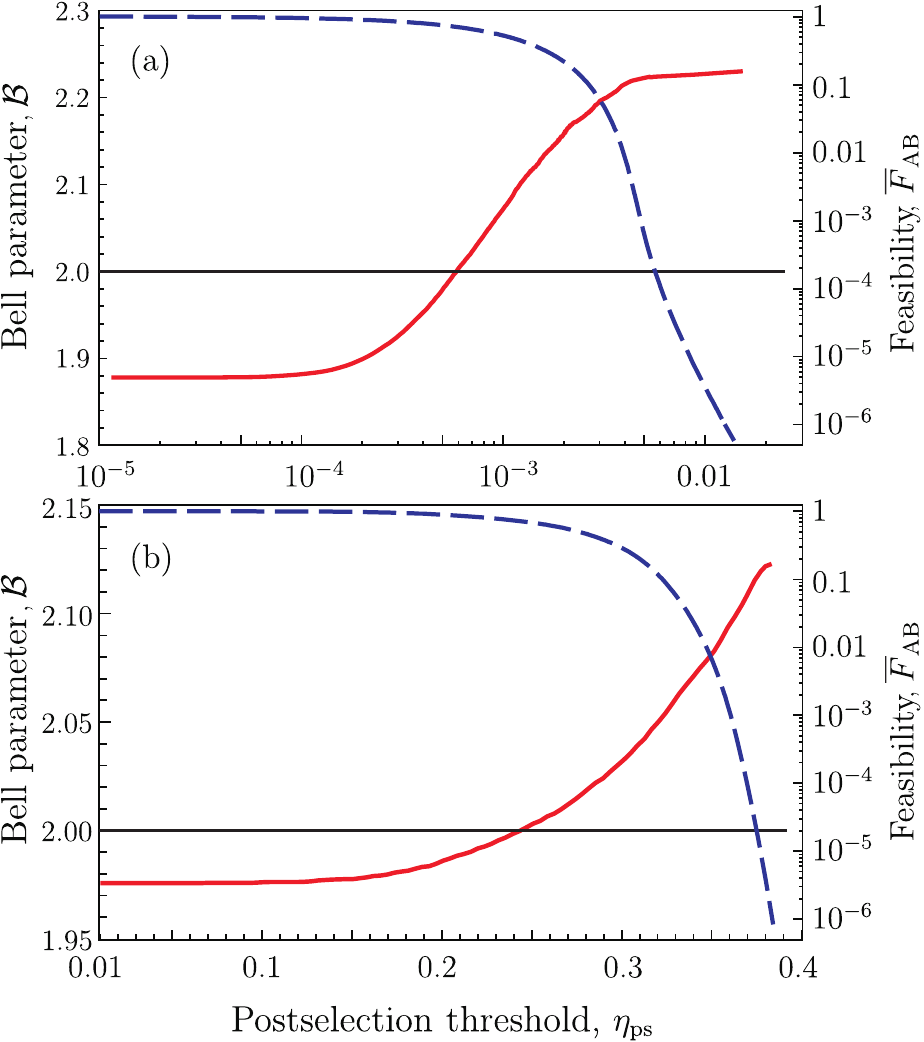}
\caption{\label{Fig5}  (Color online)  The Bell parameter, $\mathcal{B}$, vs the
postselection efficiency, $\eta_\mathrm{ps}$, (solid line), and the feasibility [cf.~Eq.~(\ref{Eq:Exceed2})] of the postselection procedure, $\overline{F}_\mathrm{AB}$ (dashed line). 
Two scenarios are considered:(a) the strong-turbulence channel 
(cf.~Appendix~\ref{App:StrongChannel}), where the mean number of stray-light and dark counts is $\nu{=}4{\times}10^{{-}4}$ and $\xi{=}0.25$, and
(b) the weak- to moderate-turbulence channel 
(cf.~Appendix~\ref{App:WeakChannel}), where  $\nu{=}2{\times}10^{{-}2}$ and the squeezing parameter is $\xi{=}0.31$. The 
detection efficiency is $\eta_c{=}0.6$.} 
\end{figure}

In Fig.~\ref{Fig5} we represent the dependence of the Bell parameter on the postselection 
efficiency $\eta_\mathrm{ps}$. Under the considered conditions  verification of the Bell-inequality violations 
is impossible without the postselection procedure due to the large dark-count 
and stray-light noise, characterized by the value of $\nu$. The postselection 
procedure certainly improves the situation.

\section{Summary and Conclusions}
\label{SummaryAndConclusions}

To conclude, we note that the incorporation of double-click events and the absence 
of correlations in the transmittances are certainly destructive factors for verifications of 
Bell-inequality violations in a turbulent atmosphere. In this paper we have theoretically studied the 
corresponding experiment with two types of channels: a 1.6-km channel with  
weak- to moderate-turbulent conditions and a 144-km channel with  strong turbulence. 
We have found that the incorporation of double-click events 
does not destroy the advantages of fading channels in the scenario of copropagation. 
However, in the case of counterpropagation, with uncorrelated channels, the advantages cannot be utilized without additional procedures.

Resources of atmospheric turbulence in the counterpropagation scenario can be used with an additional 
postselection procedure studied above. This procedure appears to be feasible for both types of  
considered channels, i.e., for weak- to moderate and strong turbulence. We believe that our results will be useful for study of quantum 
communication through atmospheric channels.

\acknowledgements
A.A.S., D.V., and W.V. acknowledge support from the Deutsche Forschungsgemeinschaft  
through Project No VO 501/21-1.

\appendix

\section{Weak- to moderate-turbulence channel}
\label{App:WeakChannel}

In this Appendix we remind the reader of the main result in Ref.~\cite{Vasylyev2016} for 
the PDT in the elliptic-beam approximation with parameters appropriate for the 
weak- to moderate-turbulence channel. In this case, the PDT is given by
\begin{align}
\mathcal{P}\left(\eta\right){=}\frac{2}{\pi}\int_{\mathbb{R}^4}\D^4\mathbf{v}
\int\limits_{0}^{ \pi /2}\D\chi \,
\rho_G(\mathbf{v};\boldsymbol{\mu},\Sigma)
\delta\left[\eta{-}\eta_\mathrm{m}\eta\left(\mathbf{v},\chi\right)\right].\label{PDTC_ElipticBeam}
\end{align}
Here $\mathbf{v}=\big(x_0\,y_0\,\Theta_{1}\,\Theta_{2}\big)^\mathrm{T}$ is a random 
vector, where $x_0$, $y_0$ are the beam-centroid coordinates, $\Theta_{1}$, 
$\Theta_{2}$ are parameters characterizing the elliptic deformations of the beam,
$\rho_G(\mathbf{v};\boldsymbol{\mu},\Sigma)$ is the Gaussian probability density of  
vector $\mathbf{v}$ with mean  $\boldsymbol{\mu}$ and covariance matrix 
$\Sigma$, $\chi$ is a uniformly distributed angle characterizing the direction of the beam-spot 
ellipse, and $\eta_\mathrm{m}$ is the efficiency related to the deterministic losses that 
occur in the channel. The transmittance $\eta\left(\mathbf{v},\chi\right)$ as a function of 
these parameters reads as
\begin{align}
&\eta\left(\mathbf{v},\chi\right)=\eta_{0}\left(\Theta_1,\Theta_2\right) \label{Tapprox}\\
&\times\exp\left\{-\left[\frac{r_0/a}
{R\left(\frac{2}{W_{\rm
			eff}\left(\Theta_1,\Theta_2,\chi\right)}\right)}\right]^{\lambda\bigl(\frac{2}{W_{\rm
			eff}\left(\Theta_1,\Theta_2,\chi\right)}\bigr)}\right\}.\nonumber
\end{align}
In this equation $r_0=\sqrt{x_0^2+y_0^2}$ is the distance between the beam and the aperture 
centers, $a$ is the radius of the receiver aperture,
\begin{align}
&W_\textrm{eff}^2\left(\Theta_1,\Theta_2,\chi\right){=}4a^2
\Bigl[\mathcal{W}\Bigl(\frac{4a^2}{
	W_1\left(\Theta_1\right)W_2\left(\Theta_2\right)}\nonumber\\
&\times e^
{\frac{a^2}{W_1^2\left(\Theta_1\right)}\bigl\{1+2\cos^2\!\chi\bigr\}}
e^
{\frac{a^2}{W_2^2\left(\Theta_2\right)}\bigl\{1+2\sin^2\!\chi\bigr\}}\Bigl)\Bigr]^{-1},\label{Weff}
\end{align}
\begin{align}
&\eta_{0}\left(\Theta_1,\Theta_2\right)\nonumber\\
&{=}1{-}
\I_0\Bigl(a^2\Bigl[\frac{1}{W_1^2\left(\Theta_1\right)}{-}\frac{1}{W_2^2\left(\Theta_2\right)}
\Bigr]\Bigr)
e^{-a^2\bigl[\frac{1}{W_1^2\left(\Theta_1\right)}{+}
	\frac{1}{W_2^2\left(\Theta_2\right)}\bigr]}\nonumber\\
&{-}2\left[1{-}e^{-\frac{a^2}{2}\!
	\bigl(\frac{1}{W_1\left(\Theta_1\right)}{-}\frac{1}{W_2\left(\Theta_2\right)}\bigr)^{2}}\!\right]\nonumber\\
&\times\exp\!\left\{\!{-}
\Biggl[\!\frac{\frac{(W_1\left(\Theta_1\right)+
		W_2\left(\Theta_2\right))^2}{|W_1^2\left(\Theta_1\right)-W_2^2\left(\Theta_2\right)|}}
{R\left(\frac{1}{W_1\left(\Theta_1\right)}{-}\frac{1}{W_2\left(\Theta_2\right)}\right)}\!\Biggr]
^{\!\lambda\left(\!\frac{1}{W_1\left(\Theta_1\right)}{-}\frac{1}{W_2\left(\Theta_2\right)}\right)}\right\},
\end{align}
$R(\xi)$ and $\lambda(\xi)$ are  scale and shape functions, respectively,
\begin{align}
R\left(\xi\right)=\Bigl[\ln\Bigl(2\frac{1-\exp[-\frac{1}{2} a^2
	\xi^2]}{1-\exp[-a^2\xi^2]\I_0\bigl(a^2\xi^2\bigr)}\Bigr)\Bigr]^{-\frac{1}{
		\lambda(\xi)}},
\end{align}
\begin{align}
\lambda\left(\xi\right)&=2a^2\xi^2\frac{e^{-a^2\xi^2}\I_1(a^2\xi^2)}{1-\exp[
	-a^2\xi^2 ] \I_0\bigl(a^2\xi^2\bigr)}\nonumber\\
&{\times}\Bigl[\ln\Bigl(2\frac{1-\exp[-\frac{1}{2} a^2 
	\xi^2]}{1-\exp[-a^2\xi^2]\I_0\bigl(a^2\xi^2\bigr)}\Bigr)\Bigr]^{-1},
\end{align}
\begin{align}
W_{i}^2{=}W_0^2\exp\Theta_{i},\qquad i=1,2,
\end{align}
$W_0$ is the beam-spot radius at the source, $\I_{i}(\xi)$ is the modified Bessel 
function of the $i$-th order, and $\mathcal{W}(\xi)$ is the 
Lambert $W$ function \cite{Corless}.

The elements of the covariance matrix $\Sigma$ and the vector of mean values 
$\boldsymbol{\mu}$ can be written in terms of the field correlation functions of the 
second and fourth orders. For conditions of weak to moderate turbulence when 
the Kolmogorov turbulence spectrum is applicable, the 
corresponding non-zero elements are given by
\begin{align}
\left\langle \Theta_{1/2}\right\rangle=\ln\Biggl[\frac{\left(1+2.96 
	\sigma_R^2\Omega^{\frac{5}{6}}\right)^2}{\Omega^2\sqrt{\left(1+2.96 
		\sigma_R^2\Omega^{\frac{5}{6}}\right)^2+1.2\sigma_R^2\Omega^{\frac{5}{6}}}}\Biggr],
\label{Eq:Theta}
\end{align}
\begin{align}
\left\langle\Delta x_0^2\right\rangle=\left\langle\Delta y_0^2\right\rangle=
0.33\,W_0^2 \sigma_R^2 \Omega^{-\frac{7}{6}},\label{Eq:x02}
\end{align}
\begin{align}
\left\langle \Delta \Theta_{1/2}^2\right\rangle=\ln\Biggl[1+\frac{1.2\sigma_R^2
	\Omega^{\frac{5}{6}}}{\left(1+2.96\sigma_R^2\Omega^{\frac{5}{6}}\right)^2}\Biggr],
\label{Eq:Theta2}
\end{align}
\begin{align}
\left\langle \Delta \Theta_1\Delta \Theta_2\right\rangle=\ln\Biggl[1-\frac{0.8\sigma_R^2
	\Omega^{\frac{5}{6}}}{\left(1+2.96\sigma_R^2\Omega^{\frac{5}{6}}\right)^2}\Biggr],
\label{Eq:Theta1Theta2}
\end{align}
where the Rytov parameter $\sigma_R^2$ is defined in Eq.~(\ref{RytovParameter}), 
$\Omega{=}\frac{kW_0^2}{2L}$ is the Fresnel parameter, $k$ is the wave number, and $L$ is the propagation 
distance. For the considered 1.6-km channel (cf.~Ref.~\cite{Usenko}), the 
corresponding parameters are 
$\sigma_R^2{=}1.5$, $\Omega{=}0.98$, 
$W_0{=}0.02\, [\mathrm{m}]$, $a{=}0.04\, [\mathrm{m}]$, and $L{=}1.6\,[\mathrm{km}]$. 
The corresponding elements of the covariance matrix $\Sigma$ and the vector of mean values 
$\boldsymbol{\mu}$ can be calculated explicitly,
\begin{align}
\Sigma=\left(\begin{array}{cccc}
2{\times} 10^{{-}4}\,[\mathrm{m^2 }]&0&0&0\\
0&2{\times} 10^{{-}4}\,[\mathrm{m^2 }]&0&0\\
0&0&0.06&{-}0.04\\
0&0&{-}0.04&0.06
\end{array}\right),
\end{align}
\begin{align}
\boldsymbol{\mu}=
\big(0\,0\,\,1.69\,\,1.69\big)^\mathrm{T},
\end{align}
which makes it possible to perform the integration in Eq.~(\ref{PDTC_ElipticBeam}) 
numerically.

The numerical integration can be performed within the Monte Carlo method. 
In this paper we have to calculate the means of a certain 
function of transmittance, $\langle f(\eta)\rangle$. For this purpose we should simulate the
$N$ values of the vector $\mathbf{v}$ and the angle $\chi$.  
The needed quantity is then estimated as
\begin{equation}
\langle 
f(\eta)\rangle=\frac{1}{N}\sum\limits_{i=1}^Nf(\eta_\mathrm{m}\eta\left(\mathbf{v}_i,\chi_i\right)),
\end{equation}
where $\eta\left(\mathbf{v}_i,\chi_i\right)$ is obtained from Eq.~(\ref{Tapprox}).  The 
deterministic losses associated with absorption and scattering by the atmosphere 
and 
optical elements correspond to the efficiency $\eta_\mathrm{m}=0.75$.

\section{Strong-turbulence channel}
\label{App:StrongChannel}

In this Appendix we consider how to estimate the PDT from the experimental data in 
\cite{Capraro} for the 144\,km channel on the Canary Islands. The Rytov parameter
for this channel satisfies the  condition $\sigma_R^2{\gg}1$, and therefore the optical 
turbulence in the channel is strong. As shown in 
\cite{Vasylyev2016}, for strong-turbulence conditions one can also use the elliptic-beam 
approximation considered in Appendix~\ref{App:WeakChannel}. However, in this case  
Eqs.~(\ref{Eq:Theta}), (\ref{Eq:x02}), (\ref{Eq:Theta2}), and (\ref{Eq:Theta1Theta2}) are not 
valid. 

It has been demonstrated both theoretically (cf.~Ref.~\cite{Vasylyev2016}) and 
experimentally (cf.~Ref.~\cite{Capraro}) that for such conditions the PDT can be 
approximated by the truncated log-normal distribution,
\begin{align}
\mathcal{P}(\eta)=\frac{1}{\eta 
	F(\eta_\mathrm{m})\sigma\sqrt{2\pi}}\exp\left[-\frac{\Bigl(\ln\eta+\mu\Bigr)^2}{2\sigma^2}\right]
\end{align}
for $\eta\in[0,\eta_\mathrm{m}]$ and 0 elsewhere, where $\sigma$ and $\mu$ are 
parameters of this distribution, and $F(\eta_\mathrm{m})$ is 
the cumulative probability distribution of the (nontruncated) log-normal distribution  
at point $\eta=\eta_\mathrm{m}$, which corresponds to deterministic losses in the channel.

From Ref.~\cite{Capraro} we already know 
important parameters: the mean radiation energy (in $\hbar\omega$ units), $\langle q 
\rangle{=}234$, in the counting time interval, its 
standard deviation $\langle \Delta q^2 \rangle{=}349^2$, and the mean 
losses $\langle \eta\rangle = 10^{-3}$. By using the relation
\begin{align}
\frac{\langle \Delta\eta^2\rangle}{\langle \eta\rangle^2}=
\frac{\langle \Delta q^2\rangle}{\langle q\rangle^2},
\end{align} 
one gets $\langle \Delta\eta^2\rangle{=}2.2{\times}10^{{-}6}$. The parameters $\sigma$ 
and $\mu$ can now be approximately evaluated as
\begin{align}
\sigma^2\approx\ln\left(1+\frac{\langle\Delta\eta^2\rangle}{\langle\eta\rangle^2}\right),\label{SigmaLogNormal1}
\end{align}
\begin{align}
\mu\approx-\ln\left(\frac{\langle\eta\rangle}{\sqrt{1+\frac{\langle\Delta\eta^2\rangle}{\langle 
			\eta\rangle^2}}}\right).
\end{align}
For the considered channel these parameters are $\sigma{=}1.08$, 
$\mu=7.49$. The deterministic losses, 
0.1\,dB/km (cf.~Ref.~\cite{DetLosses}) as well as the losses related to the optical elements 
correspond to the efficiency $\eta_\mathrm{m}{=}0.04$.

\section{Photocounting probabilities for the PDC source}
\label{App:Coefficients}

In this Appendix we discuss the probabilities 
$P_\mathrm{same}\left(\theta_\mathrm{A},\theta_\mathrm{B}\right)$
and $P_\mathrm{different}\left(\theta_\mathrm{A},\theta_\mathrm{B}\right)$ for
the case of a PDC source, cf.~Eq.~(\ref{ProbabilitySpecialSame2}). For convenience we use the Galuber-Sudarshan
$P$-representation~\cite{Glauber,GlauberPRA,Sudarshan}. The $P$ function
of the PDC state, (\ref{PDC1}), cannot be represented in terms of regular
functions. For this reason we use the corresponding characteristic function,
\begin{align}
 \Phi\left(\boldsymbol{\beta}\right)=
 \Tr\left[\exp\big(\boldsymbol{\beta}\hat{\boldsymbol{a}}^\dagger\big)
 \exp\big(-\boldsymbol{\beta}^\ast\hat{\boldsymbol{a}}\big)\hat{\rho}\right],
\end{align}
where $\hat{\boldsymbol{a}}$ is the vector consisting of field-mode annihilation
operators, $\boldsymbol{\beta}$ is the corresponding complex-number vector of the
characteristic-function arguments, and $\hat{\rho}$ is the density operator.

In this representation Eq.~(\ref{ProbabilitySquash}) reads as 
\begin{align}
P_{i_\mathrm{A}, i_\mathrm{B}}\left(\theta_\mathrm{A},
\theta_\mathrm{B}\right)=\label{Eq:ProbCharF}\\
\int\limits_{-\infty}^{+\infty}d^{8}\boldsymbol{\beta}\,
\Phi_\mathrm{out}\left(\boldsymbol{\beta}\right)
&\big[K_C\left(\beta_{i_\mathrm{A}}\right)K_C\left(\beta_{i_\mathrm{B}}\right)
K_0\left(\beta_{j_\mathrm{A}}\right)K_0\left(\beta_{j_\mathrm{B}}\right)\nonumber&
\\
+&K_C\left(\beta_{i_\mathrm{A}}\right)K_C\left(\beta_{i_\mathrm{B}}\right)
K_0\left(\beta_{j_\mathrm{A}}\right)K_C\left(\beta_{j_\mathrm{B}}\right)\nonumber&
\\
+&K_C\left(\beta_{i_\mathrm{A}}\right)K_C\left(\beta_{i_\mathrm{B}}\right)
K_C\left(\beta_{j_\mathrm{A}}\right)K_0\left(\beta_{j_\mathrm{B}}\right)\nonumber&\\
+&K_C\left(\beta_{i_\mathrm{A}}\right)K_C\left(\beta_{i_\mathrm{B}}\right)
K_C\left(\beta_{j_\mathrm{A}}\right)K_C\left(\beta_{j_\mathrm{B}}\right)\big]\nonumber.&
\end{align}
Here $\Phi_\mathrm{out}\left(\boldsymbol{\beta}\right)$ is the characteristic function of
the state
after passing the atmosphere,
$d^{8}\boldsymbol{\beta}=
d^{2}\beta_{T_\mathrm{A}}d^{2}\beta_{R_\mathrm{A}}d^{2}\beta_{T_\mathrm{B}}d^{2}
\beta_{R_\mathrm{B}}$,
\begin{align}
&K_0\left(\beta\right)=\frac{1}{\pi\eta_c}\exp\left[-\frac{|\beta|^2}{\eta_c}-\nu\right],\\
&K_C\left(\beta\right)=\delta\left(\beta\right)-K_0\left(\beta\right).
\end{align}
For more details on the notations, see the explanations following Eq.~(\ref{POVM1}).

 The output modes of the polarization analyzers,
 $\boldsymbol{\beta}{=}\{\beta_{T_\mathrm{A}},\beta_{R_\mathrm{A}},
 \beta_{T_\mathrm{B}},\beta_{R_\mathrm{B}}\}$, are related to
 the corresponding input modes,
 $\boldsymbol{\beta}{=}\{\beta_\mathrm{H_A},\beta_\mathrm{V_A},
 \beta_\mathrm{H_B},\beta_\mathrm{V_B}\}$, via the
 input-output relations,
 \begin{align}
&\beta_\mathrm{H_{A(B)}}=\beta_{T_\mathrm{A(B)}}\cos\theta_\mathrm{A(B)}-
 \beta_{R_{A(B)}}
 \sin\theta_\mathrm{A(B)}\label{IOR1},\\
 &\beta_\mathrm{V_{A(B)}}=\beta_{T_\mathrm{A(B)}}\sin\theta_\mathrm{A(B)}+
 \beta_{R_\mathrm{A(B)}}\cos\theta_\mathrm{A(B)}\label{IOR2}
 \end{align}
[cf. also the corresponding operator form given by Eqs.~(\ref{IORop1}) and
 (\ref{IORop2})]. Here the indexes $\mathrm{H_{A(B)}}$ and $\mathrm{V_{A(B)}}$ denote
 the
corresponding horizontal and vertical polarization modes, respectively. The characteristic
function of the light passed through the receiver apertures,
$\Phi_\mathrm{out}\left(\beta_\mathrm{H_A},\beta_\mathrm{V_A},
\beta_\mathrm{H_B},\beta_\mathrm{V_B}\right)$, is expressed in terms of the
characteristic function of the light generated by the source,
$\Phi_\mathrm{in}\left(\beta_\mathrm{H_A},\beta_\mathrm{V_A},
\beta_\mathrm{H_B},\beta_\mathrm{V_B}\right)$, according to the quantum-state
input-output relation for fading channels (cf.~Refs.~\cite{Semenov2009,Vasylyev2012}),
 \begin{align}
 	&\Phi_\mathrm{out}\left(\beta_\mathrm{H_A},\beta_\mathrm{V_A},
 	\beta_\mathrm{H_B},\beta_\mathrm{V_B}\right)\label{Eq:IORChF}\\
 	&=\Big\langle
 	\Phi_\mathrm{in}\left(\sqrt{\eta_\mathrm{A}}\beta_\mathrm{H_A},
 	\sqrt{\eta_\mathrm{A}}\beta_\mathrm{V_A},
 	\sqrt{\eta_\mathrm{B}}\beta_\mathrm{H_B},\sqrt{\eta_\mathrm{B}}\beta_\mathrm{V_B}\right) \Big\rangle,
 	\nonumber
 \end{align}
 where the averaging with the channel transmissions is defined by Eq.~(\ref{Eq:mean}).

The characteristic function of the PDC state, cf.~Eq.~(\ref{PDC1}), is given by
\begin{widetext}
	\begin{eqnarray}
	\Phi_\mathrm{in}\left(\beta_\mathrm{H_{A}},\beta_\mathrm{V_{A}},
	\beta_\mathrm{H_{B}},\beta_\mathrm{V_{B}}\right)&=&\exp\left[-\frac{\tanh^{2}\xi
		\left|\beta_\mathrm{V_{A}}\right|^2+\tanh^{2}\xi
		\left|\beta_\mathrm{H_{B}}\right|^2-\tanh\xi\left(\beta_\mathrm{V_{A}}\beta_\mathrm{
		 H_{B}}+
		\beta^\ast_\mathrm{V_{A}}\beta^\ast_\mathrm{H_{B}}
		\right)}{1-\tanh^{2}\xi}\right]\label{CharFuncPDC}\\
	&\times& \exp\left[-\frac{\tanh^{2}\xi
		\left|\beta_\mathrm{H_{A}}\right|^2+\tanh^{2}\xi
		\left|\beta_\mathrm{V_{B}}\right|^2+\tanh\xi\left(\beta_\mathrm{H_{A}}
		\beta_\mathrm{V_{B}}+
		\beta^\ast_\mathrm{H_{A}}\beta^\ast_\mathrm{V_{B}}
		\right)}{1-\tanh^{2}\xi}\right].\nonumber
	\end{eqnarray}
	\end{widetext}
 Substituting this expression into Eq.~(\ref{Eq:IORChF}) and then utilizing it in
 Eq.~(\ref{Eq:ProbCharF}) together with Eqs.~(\ref{IOR1}) and (\ref{IOR2}), one gets 
 the explicit
 form for $P_{i_\mathrm{A}, i_\mathrm{B}}\left(\theta_\mathrm{A},
 \theta_\mathrm{B}\right)$.  Finally the result  used in Eqs.~(\ref{Eq:Psame}) and
 (\ref{Eq:Pdifferent}) which leads to the expressions for
 $P_\mathrm{same}\left(\theta_\mathrm{A},\theta_\mathrm{B}\right)$
 and $P_\mathrm{different}\left(\theta_\mathrm{A},\theta_\mathrm{B}\right)$ is given by
 Eq.~(\ref{ProbabilitySpecialSame2}).

\section{Bell states}
\label{App:Bell states}

In this Appendix we discuss the derivation of the expression for the maximal value of
the Bell parameter [cf.~ Eq.~(\ref{S_OnOff})] in the case of weak-intensity sources
described by the Bell state~(\ref{BellState1}) and (\ref{BellState2}). First, we recall 
the 
explicit form for the
density operator of the corresponding state after passing through the atmosphere 
(cf.~Ref.~\cite{Semenov2010}),
\begin{equation}
\hat{\rho}=p_{\scriptscriptstyle\mathrm{0}}
\hat{\rho}_{\scriptscriptstyle\mathrm{0}}+ p_{\scriptscriptstyle
\mathrm{H_{A}}}\hat{\rho}_{\scriptscriptstyle\mathrm{H_{A}}} +
p_{\scriptscriptstyle
\mathrm{V_{A}}}\hat{\rho}_{\scriptscriptstyle\mathrm{V_{A}}} +
p_{\scriptscriptstyle
\mathrm{H_{B}}}\hat{\rho}_{\scriptscriptstyle\mathrm{H_{B}}} +
p_{\scriptscriptstyle
\mathrm{V_{B}}}\hat{\rho}_{\scriptscriptstyle\mathrm{V_{B}}} +
p_\mathcal{\scriptscriptstyle
B}\hat{\rho}_\mathcal{\scriptscriptstyle
B},\label{BellStateTurbulence}
\end{equation}
where $\hat{\rho}_{\scriptscriptstyle\mathrm{0}}$ is the density operator of the vacuum
state, $\hat{\rho}_{\scriptscriptstyle\mathrm{H_{A(B)}}}$,
$\hat{\rho}_{\scriptscriptstyle\mathrm{V_{A(B)}}}$ are the single-photon density operators
in the corresponding mode, and $\hat{\rho}_\mathcal{\scriptscriptstyle
B}=\ket{\mathcal{B}}\bra{\mathcal{B}}$ is the density operator of the Bell state [cf.
Eq.~(\ref{BellState1})]. Here $p_{\scriptscriptstyle\mathrm{0}}$ is the probability that 
no photons from the source will reach the receivers [cf.~Eq.~(\ref{Eq:p0})], 
$p_\mathcal{\scriptscriptstyle B}$ is the
probability that both photons of the Bell state will reach the receivers 
[cf.~Eq.~(\ref{Eq:pB})],
\begin{align}
	p_{\scriptscriptstyle \mathrm{H_{A(B)}}}=p_{\scriptscriptstyle \mathrm{V_{A(B)}}}=
	\frac{1}{2}\left\langle \eta_\mathrm{A(B)}
	\left(1-\eta_\mathrm{B(A)}\right)\right\rangle
\end{align}
is the probability that only one photon in the corresponding mode reaches the 
receivers such
that
\begin{align}
	p_1=	p_{\scriptscriptstyle \mathrm{H_{A}}}+	p_{\scriptscriptstyle \mathrm{H_{B}}}+	
	p_{\scriptscriptstyle \mathrm{V_{A}}}+	p_{\scriptscriptstyle \mathrm{V_{B}}}
\end{align}
is the probability that only one photon will reaches the receivers [cf.~Eq.~(\ref{Eq:p1})].

Next we utilize the operator form of the polarization-analyzer input-output relations
[cf.~Eqs.~(\ref{IORop1}) and (\ref{IORop2})]
 in Eqs.~(\ref{POVM}) and (\ref{POVM1}). The result together with the density operator,
 (\ref{BellStateTurbulence}), is substituted in Eq.~(\ref{ProbabilitySquash}), which gives us the
 probabilities $P_{i_\mathrm{A}, i_\mathrm{B}}\left(\theta_\mathrm{A},
 \theta_\mathrm{B}\right)$. By using these probabilities in Eqs.~(\ref{Eq:Psame}),
 (\ref{Eq:Pdifferent}), and (\ref{correlation}) one gets the correlation coefficient,
\begin{align}
&E\left(\theta_\mathrm{A}, \theta_\mathrm{B}\right) =\label{Eq:CorrelationBellState}\\
&-\frac{p_{\scriptscriptstyle
		\mathcal{B}}\eta_c^2e^{2\nu}\cos\left[2(\theta_\mathrm{A}-\theta_\mathrm{B})\right]}
{p_{\scriptscriptstyle
\mathcal{B}}\left(e^{2\nu}{+}\eta_c{-}1\right)^2{+}p_{\scriptstyle\mathrm{0}}
\left(e^{2\nu}{-}1\right)^2{+}
p_{\scriptstyle 1}\left(e^{2\nu}{-}1\right)
\left(e^{2\nu}{+}\eta_c{-}1\right)}.\nonumber
\end{align}
For the particular values of the polarization angles, $\big(\theta_\mathrm{A}^{(1)},
\theta_\mathrm{B}^{(1)},\theta_\mathrm{A}^{(2)},
\theta_\mathrm{B}^{(2)}\big)=\big(0,\frac{\pi}{8},\frac{\pi}{4},\frac{3\pi}{8}\big)$, the Bell
parameter $\mathcal{B}$, cf.~Eq.~(\ref{BellParameter}), takes the value given by
Eq.~(\ref{S_OnOff}).

\section{Bell-inequality test with discarding double-click 
events}
 \label{App:NDubleClicks}

  In this Appendix we remind the reader of some results in Ref.~\cite{Semenov2010},
  where the effect of double-click events has not been included. In this case the last three
  terms in Eq.~(\ref{ProbabilitySquash}) vanish. For the PDC source the probabilities
  $P_\mathrm{same}\left(\theta_\mathrm{A},\theta_\mathrm{B}\right)$
  and $P_\mathrm{different}\left(\theta_\mathrm{A},\theta_\mathrm{B}\right)$, read as
  \begin{widetext}
  	\begin{align}
  	&P_\mathrm{i}\left(\theta_\mathrm{A},
  	\theta_\mathrm{B}\right)=
  	\nonumber\\
  	&2\left(1-\tanh^2\xi\right)^4
  	\left[\left\langle\frac{\exp\left(-2\nu\right)}
  	{C_\mathrm{0}+C_\mathrm{1A}+C_\mathrm{1B}+C_\mathrm{i}}
  	\right\rangle-
  	\left\langle\frac{\exp\left(-3\nu\right)}{C_\mathrm{0}+C_\mathrm{1A}}
  	\right\rangle
  	-
  	\left\langle\frac{\exp\left(-3\nu\right)}{C_\mathrm{0}+C_\mathrm{1B}}
  	\right\rangle
  	+\left\langle\frac{\exp\left(-4\nu\right)}{C_\mathrm{0}}\right\rangle\right],
  	\label{ProbabilitySpecialSame1}
  	\\
  	\nonumber
  	\end{align}
where $i{=}\{\mathrm{same},\mathrm{different}\}$, for coefficients see 
Eqs.~(\ref{Eq:C02})-(\ref{Eq:Cdifferent2}). For the weak-intensity source, described by the 
Bell
state [cf.~Eq~(\ref{BellState1})], the maximal value of the Bell parameter is given by
	\begin{align}
	&\mathcal{B}=\frac{2\sqrt{2}\,p_{\scriptscriptstyle
			\mathcal{B}}\eta_c^2e^{2\nu}}
	{p_{\scriptscriptstyle
			\mathcal{B}}\left[\left(1-\eta_c\right)
		\left(e^{\nu}-2\right)+e^{\nu}\right]^2+
		2p_{\scriptscriptstyle 1}\left[e^{\nu}-1\right]
		\left[\eta_c e^{\nu}+
		2\left(e^{\nu}-1\right)\left(1-\eta_c\right)\right]
		+4p_{\scriptscriptstyle\mathrm{0}}
		\left[e^{\nu}-1\right]^2},\label{S_OnOff1}
	\end{align}
for more details see the explanations following Eq.~(\ref{S_OnOff}). 	
\end{widetext}


\begin{thebibliography}{99}
\bibitem{Takesue} H. Takesue, S. W. Nam, Q.  Zhang, R.  H.  Hadfield, T. Honjo, K.
        Tamaki, and Y. Yamamoto, Quantum Key Distribution over 40 dB
        Channel Loss Using Superconducting Single Photon Detectors, Nature Photon. \textbf{1}, 343 (2007).
\bibitem{Satellite0} %airplane
        S. Nauerth, F. Moll, M. Rau, C. Fuchs, J. Horwath, S. Frick, and H. Weinfurter,
        Air-to-Ground Quantum Communication, Nat. Photon. {\bf 7}, 382  (2013).
\bibitem{Satellite2} J.-Yu Wang {\it et al.}, Direct and Full-Scale Experimental Verifications
        Towards Ground-Satellite Quantum Key Distribution, Nat. Photon.
        {\bf 7}, 387 (2013).
\bibitem{Satellite3} G. Vallone, D. Bacco, D. Dequal, S. Gaiarin, V. Luceri, G. Bianco,
        and P. Villoresi, Experimental Satellite Quantum Communications,
        Phys. Rev. Lett. {\bf 115}, 040502 (2015).
\bibitem{Satellite4} D. Dequal, G. Vallone, D. Bacco, S. Gaiarin, V. Luceri, G.
        Bianco, and P. Villoresi, Experimental Single Photon Exchange along a Space
        Link of 7000 km, Phys. Rev. A {\bf 93}, 010301 (2016).
\bibitem{Satellite5} G. Vallone, D. Dequal, M. Tomasin, F. Vedovato, M.
        Schiavon, V. Luceri, G. Bianco, and P. Villoresi,
        Quantum interference along satellite-ground channels, Phys. Rev. Lett. {\bf 116}, 253601 (2016).
\bibitem{Satellite6} J.-P.~Bourgoin {\it et al.}, A Comprehensive Design and Performance
        Analysis of Low Earth Orbit Satellite Quantum Communication, New J. Phys.
        {\bf 15}, 023006 (2013).
\bibitem{BellReview} N. Brunner, D. Cavalcanti, S. Pironio,
        V. Scarani, and S. Wehner, Bell Nonlocality, Rev. Mod. Phys. \textbf{86}, 419 (2014).
\bibitem{Ekert} A. K. Ekert, Quantum Cryptography Based on Bell's Theorem,
        Phys. Rev. Lett. \textbf{67}, 661 (1991).
\bibitem{Acin1} A.~Ac\'in, N. Gisin, and L. Masanes, From Bell’s Theorem to Secure Quantum Key Distribution,
	Phys. Rev. Lett. \text{97}, 120405 (2006).      
\bibitem{Acin2} A.~Ac\'in,  N. Brunner, N. Gisin, S. Massar, S. Pironio, and V. Scarani, 
Device-Independent	Security of Quantum Cryptography against Collective Attacks,
	Phys. Rev. Lett. \text{98}, 230501 (2007).        
\bibitem{Ursin} R. Ursin {\it et al.},
         Entanglement-Based Quantum Communication over 144 km, Nature Phys. {\bf 3}, 481 (2007).
\bibitem{Fedrizzi} A. Fedrizzi, R. Ursin, T. Herbst, M. Nespoli, R. Prevedel, T. Scheidl, F. Tiefenbacher, T.
         Jennewein, and A. Zeilinger, High-Fidelity Transmission of Entanglement over a High-Loss
         Free-Space Channel, Nature Phys. {\bf 5}, 389 (2009).
\bibitem{Semenov2009} A. A. Semenov and W. Vogel, Quantum Light in the Turbulent Atmosphere,
        Phys. Rev. A {\bf 80}, 021802(R) (2009).
\bibitem{Vasylyev2012}D. Yu. Vasylyev, A. A. Semenov and W. Vogel, Toward Global
         Quantum Communication: Beam Wandering Preserves Nonclassicality,
         Phys. Rev. Lett. \textbf{108}, 220501 (2012).
\bibitem{Avetisyan}  H. Avetisyan and C. H. Monken, Higher order correlation beams in atmosphere under strong 			turbulence conditions, Opt. Express \textbf{24}, 2318 (2016).
\bibitem{Zhang} Y. Zhang, Sh. Prabhakar, A. H. Ibrahim, F. S. Roux, A. Forbes, and Th. Konrad, Experimentally observed 		decay of high-dimensional entanglement through turbulence, Phys. Rev. A \textbf{94}, 032310 (2016).         
\bibitem{Semenov2010} A. A. Semenov and W. Vogel, Entanglement Transfer through the
         Turbulent Atmosphere, Phys. Rev. A \textbf{81}, 023835 (2010); \textbf{85}, 019908(E)
         (2012).
\bibitem{Ma} X. Ma, C. H. F. Fung, and H. K. Lo, Quantum Key Distribution with
          Entangled Photon Sources, Phys. Rev. A \textbf{76}, 012307 (2007).
\bibitem{Kok} P. Kok and S. L. Braunstein, Postselected Versus Nonpostselected
         Quantum Teleportation Using Parametric Down-Conversion, Phys. Rev. A
         \textbf{61}, 042304 (2000).
\bibitem{Semenov2011}A. A. Semenov and W. Vogel, Fake Violations of the
         Quantum Bell-Parameter Bound, Phys. Rev. A \textbf{83}, 032119 (2011);
         \textbf{85}, 049904(E) (2012).
\bibitem{Beaudry} N. J. Beaudry, T. Moroder, and N. L\"utkenhaus,
         Squashing Models for Optical Measurements in Quantum Communication,
         Phys. Rev. Lett. {\bf 101}, 093601 (2008).
\bibitem{Moroder} T. Moroder, O. G\"uhne, N. Beaudry, M. Piani,
         and N. L\"utkenhaus, Entanglement Verification with Realistic
         Measurement Devices via Squashing Operations, Phys. Rev. A {\bf 81}, 052342 (2010).
\bibitem{Fung} C. H. F. Fung, H. F. Chau, and H.-K. Lo, Universal
         Squash Model for Optical Communications Using Linear Optics and Threshold
         Detectors, Phys. Rev. A \textbf{84}, 020303(R) (2011).
\bibitem{CHSH} J. F. Clauser, M. A. Horn, A. Shimony, and R. A. Holt,
        Proposed Experiment to Test Local Hidden Variable Theories, Phys. Rev. Lett. {\bf 23}, 880 (1969).
\bibitem{Mandel} L. Mandel and E. Wolf, \textit{Optical Coherence and Quantum Optics}
         (Cambridge University Press, Cambridge, 1995).
\bibitem{Kelley} P. L. Kelley and W. H. Kleiner, Theory of Electromagnetic Field Measurement and Photoelectron     
         Counting, Phys. Rev. \textbf{136}, A316 (1964).         
\bibitem{Semenov2008} A. A. Semenov, A. V. Turchin and H. V. Gomonay,
        Detection of Quantum Light in the Presence of Noise, Phys. Rev. A
        {\bf 78}, 055803 (2008); {\bf 79}, 019902(E) (2009).
\bibitem{Glauber} R.~J.~Glauber, Photon Correlations, Phys. Rev. Lett.
        \textbf{10}, 84 (1963).
\bibitem{GlauberPRA} R.~J.~Glauber, Coherent and Incoherent States of
        the Radiation Field, Phys. Rev. A \textbf{131}, 2766 (1963).
\bibitem{Sudarshan} E.~C.~G.~Sudarshan, Equivalence of Semiclassical and
        Quantum Mechanical Descriptions of Statistical Light Beams, Phys. Rev. Lett.
        \textbf{10}, 277 (1963).
\bibitem{Tatarskii} V. Tatarskii, \textit{The Effect of the Turbulent Atmosphere
        on Wave Propagation} (Israel Program for Scientific Translations,
        Jerusalem, 1971).
\bibitem{Ishimaru} A. Ishimaru, \textit{Wave Propagation and Scattering in Random Media}
        (Academic Press, NY, 1978).
\bibitem{Andrews} L.~Andrews, R.~Phillips, and C.~Hopen, \textit{Laser Beam
	Scintillation with Applications} (SPIE Press, Washington, 2001).
\bibitem{Andrews2} L.~Andrews and R.~Phillips, \textit{Laser Beam Propagation through
	Random Media} (SPIE Press, Washington, 2005).
\bibitem{Fante1} R.~L.~Fante, Electromagnetic Beam Propagation in Turbulent
	Media, Proc. IEEE  \textbf{63}, 1669 (1975).
\bibitem{Fante2} R.~L.~Fante, Electromagnetic beam propagation in turbulent
	media: An update, Proc. IEEE  \textbf{68}, 1424 (1980).
\bibitem{Chumak2006} G. P. Berman and A. A. Chumak, Photon Distribution Function 
for 
Long-Distance Propagation of Partially Coherent Beams through the Turbulent 
Atmosphere,
Phys. Rev. A \textbf{74}, 013805 (2006).
\bibitem{Chumak2016} O. O. Chumak and R. A. Baskov, Strong Enhancing Effect of 
Correlations of Photon Trajectories on Laser Beam Scintillations, Phys. Rev. A \textbf{93}, 
033821 (2016).
\bibitem{Vasylyev2016} D. Vasylyev, A.~A. Semenov, and W. Vogel, Atmospheric Quantum
        Channels with Weak and Strong Turbulence, Phys. Rev. Lett. \textbf{117}, 090501 
        (2016).
\bibitem{Diament} P.~Diament and M.~C.~Teich, Photodetection of
	Low-Level Radiation through the Turbulent Atmosphere, J. Opt. Soc. Am. {\bf
	60}, 1489 (1970).
\bibitem{Perina} J.~Pe\v{r}ina, On the photon counting statistics of
	light passing through an inhomogeneous random medium, Czech. J. Phys.
        \textbf{22}, 1075 (1972).
\bibitem{Perina1973} J.~Pe\v{r}ina, V.~Pe\v{r}inova, M.~C.~Teich, and
        P.~Diament, Two Descriptions for the Photocounting Detection of
	Radiation Passed through a Random Medium: A Comparison for the Turbulent
	Atmosphere, Phys. Rev. A \textbf{7}, 1732 (1973).
\bibitem{Milonni} P.~Milonni, J.~Carter, Ch.~Peterson, and R.~Hughes,
        Effects of Propagation through Atmospheric Turbulence on Photon
	Statistics, J. Opt. B \textbf{6}, S742 (2004).
\bibitem{Capraro} I.~Capraro, A.~Tomaello, A.~Dall'Arche, F.~Gerlin, R.~Ursin,
        G.~Vallone, and P.~Villoresi, Impact of Turbulence in Long Range
	Quantum and Classical Communications, Phys. Rev. Lett. \textbf{109}, 200502
        (2012).
\bibitem{Usenko} V.~C.~Usenko, B.~Heim, C.~Peuntinger, C.~Wittmann, C.~Marquardt, G.~Leuchs, and
        R.~Filip, Entanglement of Gaussian States and the Applicability to Quantum Key
        Distribution over Fading Channels, New J. Phys. \textbf{14}, 093048 (2012).
 \bibitem{Corless} R.~Corless, G.~Gonnet, D.~Hare, D.~Jeffrey, and D.~Knuth,
 On the Lambert W Function, Adv. Comput. Math. \textbf{5}, 329 (1996).
\bibitem{DetLosses} V.~Scarani, H.~ Bechmann-Pasquinucci, N.~J.~Cerf, M.~ Du\v{s}ek, 
N.~L\"utkenhaus, and M.~Peev, The security of practical quantum key distribution, Rev. 
Mod. Phys. \textbf{81}, 1301 (2009).
\end{thebibliography}
\end{document}